\documentclass[sigconf,authorversion=true, bookmarks=false]{acmart}
\usepackage{rotating}

\newcommand{\RQOne}{How well can each chat activity metric predict developer productivity?}
\newcommand{\RQTwo}{How well can multiple chat activity metrics predict developer productivity?}

\setcopyright{acmcopyright}\acmConference[ICSEW'20]{IEEE/ACM 42nd International Conference on Software Engineering Workshops }{May 23--29, 2020}{Seoul, Republic of Korea}
\acmBooktitle{IEEE/ACM 42nd International Conference on Software Engineering Workshops (ICSEW'20), May 23--29, 2020, Seoul, Republic of Korea}

\begin{document}

\title{Chat activity is a better predictor than chat sentiment on software developers productivity}

\author{Miikka Kuutila}
\email{miikka.kuutila@oulu.fi}
\affiliation{%
  \institution{University of Oulu}
}

\author{Mika V. M{\"a}ntyl{\"a}}
\email{mika.mantyla@oulu.fi}
\affiliation{%
  \institution{University of Oulu}
}

\author{Ma{\"e}lick Claes}
\email{maelick.claes@oulu.fi}
\affiliation{%
  \institution{University of Oulu}
}
\renewcommand{\shortauthors}{Kuutila, et al.}

\begin{abstract}
Recent works have proposed that software developers' positive emotion has a positive impact on software developers' productivity. In this paper we investigate two data sources: developers chat messages (from Slack and Hipchat) and source code commits of a single co-located Agile team over 200 working days. Our regression analysis shows that the number of chat messages is the best predictor and predicts productivity measured both in the number of commits and lines of code with $R^2$ of 0.33 and 0.27 respectively. We then add sentiment analysis variables until AIC of our model no longer improves and gets $R^2$ values of 0.37 (commits) and 0.30 (lines of code). Thus, analyzing chat sentiment improves productivity prediction over chat activity alone but the difference is not massive. This work supports the idea that emotional state and productivity are linked in software development. We find that three positive sentiment metrics, but surprisingly also one negative sentiment metric is associated with higher productivity. 
\end{abstract}



\maketitle

\section{Introduction}\label{sec:introduction}

In recent times, there has been an increasing interest towards affect of software developers, and calls for studies in the field of \emph{behavioural software engineering}~\cite{lenberg2015behavioral}, and \emph{psychoempirical software engineering}~\cite{graziotin2015understanding}. Particularly, previous studies have examined happiness and unhappiness of software developers in relation to their productivity and problem solving skills~\cite{graziotin2014happy, graziotin2017unhappy}. These studied have found that in general positive emotional state are associated with higher productivity in software engineering.

While a lot of work related to sentiment analysis in software engineering has focused on data available software repositories, such as issue trackers and Q\&A websites, sentiment analysis of instant or chat messaging between software developers is scarce. This is surprising as chat due to its interactive and non-documentary purpose seems like the likeliest candidate to contain expressions of sentiment and emotion. Tools related to chat and instant messaging for software developers are also commonly mentioned studies related to global software development (gsd) and distributed development tools and processes~\cite{lanubile2010collaboration, portillo2012tools}. Yet, papers focusing on chat usage are scarce. 

Some previous work of chat usage in software engineering exists but for different purposes than productivity and sentiment analysis. 
Alkadhi et al.~\cite{alkadhi2017rationale} analyzed chat messages for rationale, defined as messages containing discussions about issues, alternatives and argumentation leading to the decisions related to software development. In their study, only 9\% of messages contained rationale, but they were able to filter out messages without rationale with precision and recall above 0.9. Chatterjee et al.~\cite{chatterjee2019exploratory} have studied Q\&A, "Question and Answer", communications in Slack chat environment. They demonstrate that that using repository mining and machine learning techniques, it is possible to extract these questions and answer conversations to produce conversations similar to those found in StackOverflow\footnote{\url{https://stackoverflow.com/}}. Instant messaging of software developers over the IRC protocol~\cite{oikarinen1993internet} have been studied by Shibab et al.~\cite{shihab2009studying}, observing how meetings are in open source software development projects and giving some insights on the contents discussions.

Metrics related to social interaction between software developers used in prediction models have been gathered in a mapping study by Wiese et al. ~\cite{wiese2014social}. One of the metrics identified from prior literature is the number of messages sent. In our prior work, we have linked number of chat messages sent to affective states and well-being~\cite{kuutila2018using}.

In this paper, we look at the relationships between software developer commit activity, chat activity and the sentiments expressed in instant messaging systems in a single co-located team. More specifically, we predict the commit activity of individual developers using various measures of their chat activity and answer the following two research questions:

\begin{description}
\item[RQ1] {\RQOne}
\item[RQ2] {\RQTwo}
\end{description}

The rest of the paper is structured as follows. The methodology for mining productivity metrics and using sentiment analysis on developer chat logs is explained in Section~\ref{sec:methodology}. In Section~\ref{sec:results} we provide the answers to our research questions. Lastly, conclusions are provided in Section~\ref{sec:conclusions}.

\section{Methodology}\label{sec:methodology}

\paragraph{Data extraction}

For the purposes of this study, we were provided access to a single software project's Git repository, and to a JSON dump of the chat room used by developers from a local Finnish company. All of the developers working on the project are paid, and usually worked in the same location during office hours. The specific tool used for communication changed during our study from Hipchat\footnote{\url{http://www.businessinsider.com/atlassian-launches-hipchat-successor-stride-2017-9}} to Slack\footnote{\url{https://slack.com/}}. The observed period is around eight months, and contains 200 working days, 7976 commits, over a million code lines changed and 30704 chat messages sent from eight different developers. We used data only from working days, from Monday to Friday.

\paragraph{Commit activity metrics}

For the purposes of this study, we extracted the list of commits from the Git repository using Perceval~\cite{duenas2018perceval}. For each chat participant we computed the number of commits made (\emph{ncommits}) and the number of lines of code modified (\emph{nloc}). These are widely used as proxy measures for productivity in prior literature~\cite{mockus2002two, boehm1981software}. The developers were instructed to commit small commits often by the project managers, and thus had multiple commits per day. This is why we decided to include commits in the analysis.

\paragraph{Sentiment analysis of chat messages}

Because the chat messages used in the analyzed project are written in Finnish, we only had access to elementary sentiment analysis techniques. We translated lexicons used in software engineering context for measuring arousal~\citep{mantyla2017bootstrapping} and valence to Finnish. The valence lexicon has been constructed in the same way as the arousal lexicon. Chat logs were lemmatized using the open source software Voikko~\citep{Voikko}, and then scored on valence and arousal using the translated lexicons. For this study, we centered the values of valence and arousal to 0, hence low valence and arousal scores are shown in values under 0 and high valence and arousal in scores over 0. We use this information in variables \emph{sent low valence \%}, \emph{sent high valence \%}, \emph{sent low arousal \%} and \emph{sent high arousal \%}. Hence, variable \emph{sent low valence \%} contains the percentage of messages containing at least one word with a valence score below 0 and the variable \emph{sent high valence \%} the percentage of messages containing at least one word with a valence score above 0. The same holds true for variables \emph{sent low arousal \%} and \emph{sent high arousal \%}. Similarly, we also computed the maximum and minimum arousal and valence scores for each day for each developer, and these are found in the variables \emph{sent minimum valence}, \emph{sent maximum valence}, \emph{sent minimum arousal} and \emph{sent maximum arousal}. During a day in which only high valence or arousal words were used, the minimum would be a positive number.

\paragraph{Emoticons and emojis in chat messages}

We extracted emoticons and emojis that were used in the chat messages. Emoticons are textual representations of human emotion using only keyboard characters such as letters, number or punctuation marks. Emojis refer to "picture characters" or pictographs~\citep{miller2016blissfully}. Similar to some of the authors' previous work~\cite{claes2018use}, we manually classified the emoticons to the basic emotions of Plutchik's wheel of emotions \citep{plutchik1991emotions}: joy, sadness, surprise, confusion and anger. From these, we calculated the percentage of messages containing \emph{emoticons} (and emojis), the percentage of messages containing emoticons and emojis \emph{related to joy} and the percentage of messages containing emoticons and emojis \emph{related to surprise, sadness, or confusion}. Due to the low number of emoticons and emojis for the latter group of emotions, we combined them in one variable named \emph{sadconfusionsurprise}. We provide the translated lexicons, the list of emoticons used and their classification in a GitHub repository\footnote{\url{https://github.com/M3SOulu/semotion2020}}.

\paragraph{Prediction models}

We answer RQ1 by building different regression models using R\footnote{\url{https://www.rdocumentation.org/packages/stats/versions/3.6.2/topics/glm}}. Each model uses as predictors a single chat metrics and a weekday variable as dummies to control for possible weekly seasonality. We evaluate them using $R^2$.

For RQ2, we build a multivariate regression model iteratively using a step-up approach~\citep{yamashita2007stepwise} evaluated using Akaike Information Criterion (AIC)~\cite{akaike1998information}. Starting from the model with the dummy weekly control variables, we add chat variables as predictors by selecting the one that produces the lowest AIC when added to the model. The iterative process stops once the AIC does not decrease.
\section{Results}\label{sec:results}

\begin{table}[ht]
\centering
\caption{Pairwise linear regression models, where the productivity variable is predicted. Number provided is $R^2$, and the sign signifies the sign of the coefficient. All predictors significant at p < 0.05 level.}
\label{tab:pw}
\begin{tabular}{rlll}
  \hline
 &  & ncommitslog & nloclog  \\ 
  \hline
1 & nchatlog & \textbf{(+)0.330} & \textbf{(+)0.268}   \\ 
  2 & emoticon \% & \textbf{(+)0.038} & \textbf{(+)0.036}   \\ 
  3 & emot joy \% & \textbf{(+)0.117} & \textbf{(+)0.098}  \\ 
  4 & emot sadconfusionsurprise \% & \textbf{(+)0.092} & \textbf{(+)0.062}   \\ 
  5 &  sent high valence \% & \textbf{(+)0.034} & \textbf{(+)0.039}   \\ 
  6 & sent low valence \% & \textbf{(+)0.043} & \textbf{(+)0.042}  \\ 
  7 & sent min valence & \textbf{(-)0.147} & \textbf{(-)0.108}   \\ 
  8 & sent max valence & \textbf{(+)0.127} & \textbf{(+)0.103}  \\ 
  9 & sent high arousal \% & \textbf{(+)0.100} & \textbf{(+)0.100}  \\ 
  10 & sent low arousal \% & \textbf{(+)0.143} & \textbf{(+)0.130}  \\ 
  11 & sent minimum arousal & \textbf{(-)0.196} & \textbf{(-)0.158}  \\ 
  12 & sent maximum arousal & \textbf{(+)0.175} & \textbf{(+)0.146}  \\ 
   \hline
\end{tabular}
\end{table}

\begin{table}[ht]
\centering
\caption{Step-wise step-up build of multivariate regression model predicting number of commits. P-value of the last added variable.}
\label{tab:swncom}
\begin{tabular}{rlllll}
  \hline
 & Added variable & AIC & $R^2$ & p-value \\ 
  \hline
  & weekdays & 5030.2 & 0.003 & - \\ 
 & nchatlog & 4396 & 0.330 &< 2e-16 \\ 
 & emot sadconfusionsurprise \% & 4353.5 & 0.348 & 2.88e-11   \\ 
 & sent max valence & 4328.1 & 0.360 & 1.75e-07   \\ 
 & sent low arousal \% & 4315.5 & 0.365 &  0.000138   \\ 
 & sent high valence \% & 4312.2 & 0.367 & 0.022 \\ 
 & emot joy \% & 4310.5 & 0.369 & 0.054 \\
   \hline
\end{tabular}
\end{table}

\begin{table}[ht]
\centering
\caption{Step-wise step-up build of multivariate regression model predicting number of lines changed. P-value of the last added variable.}
\label{tab:swnloc}
\begin{tabular}{rlllll}
  \hline
 & Added variable & AIC & $R^2$ & p-value \\ 
  \hline
 & weekdays & 8339.8 & 0.001 & \\ 
 & nchatlog & 7843.7 & 0.268 & < 2e-16 \\ 
 & sent max valence & 7822.8 & 0.279 & 1.83e-06 \\ 
 & sent high valence \% & 7801.7 & 0.289 & 1.60e-06 \\ 
 & emot sadconfusionsurprise \% & 7785.9 & 0.297 & 2.60e-05 \\ 
 & sent low arousal \% & 7778.4 & 0.301 & 0.002 \\ 
 & emot joy \% & 7778.1 & 0.302 & 0.133\\ 
   \hline
\end{tabular}
\end{table}

\paragraph{RQ1 - \RQOne}
Table~\ref{tab:pw} shows $R^2$ values for models predicting productivity measures, number of commits and number of lines of code changed. All models are controlled for weekly seasonality by introducing binary weekday variables to the regression equation. In both cases, the highest $R^2$ values are achieved by number of chat messages sent. Other predictors achieving an $R^2$ for prediction of number of commits of over 0.1 are the percentage of joy emoticons, minimum and maximum valence, as well as all variables related to arousal scores. For the prediction of number of lines of code, the other predictors achieving an $R^2$ of over 0.1 are minimum and maximum valence, as well as all variables related to arousal

\paragraph{RQ2 - \RQTwo}.
Table~\ref{tab:swncom} shows the a multivariate model predicting number of commits with variables related to chat activity and sentiment. After number of chat messages, the next predictor added to the model was the percentage of emoticons related to sadness, confusion and surprise present in messages. Variables added after that are maximum valence, percentage of low arousal words, percentage of high valence words, and the last one variable to lower the AIC-value is the percentage of joy emoticons.

Table~\ref{tab:swnloc} shows the a multivariate model predicting number of lines of code changed with variables related to chat activity and sentiment. The model produced is very similar to the one predicting commits, they both hold the same variables but in slightly different order. 

To summarize three positive emotion metrics  (sent max valence, sent high valence \%, emot joy\%) are associated with increased productivity. However, a negative emotion metric (emot sadconfusionsurprise \%)  is also associated with increased productivity. Thus the former results support the idea that positive emotion is associated with higher productivity. The latter result adds the element that also negative emotion can be associated with more productivity. We could speculate that as long as there is emotion whether positive or negative then there is productivity. Finally, we were surprised to find that when a higher percentage of messages express low arousal then productivity is higher. Perhaps, low arousal messages express confidence that is associated with productivity, for instance, "Relax I'll fix it".   

\begin{figure*}[ht]
\centering
\includegraphics[width=\textwidth]{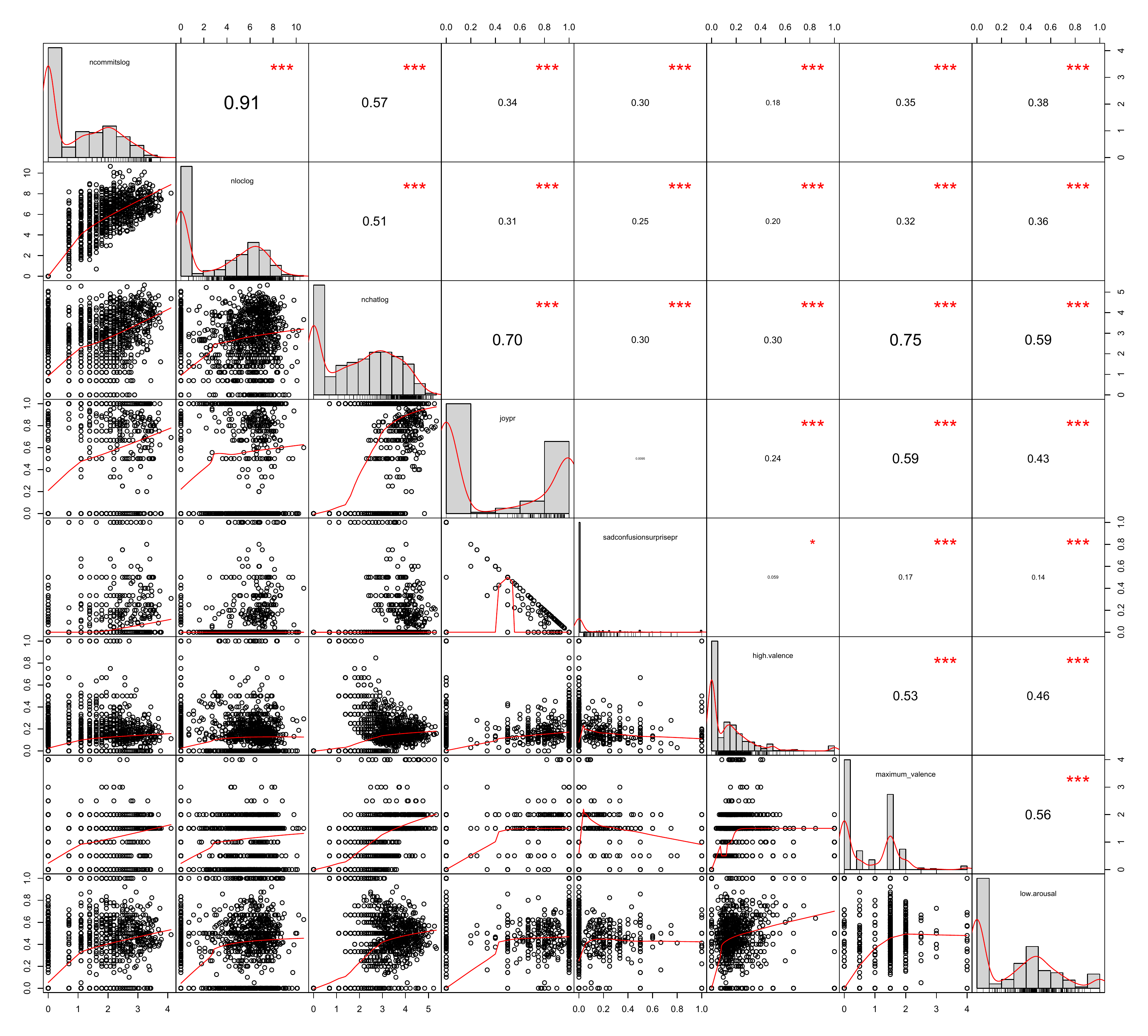}
\caption{Correlation matrix of all variables identified in step-wise builds of multiple regression models shown in Tables \ref{tab:swncom} and \ref{tab:swnloc}.}
\label{fig:image}
\end{figure*}

Lastly, we have gathered all variables presented in tables \ref{tab:swncom} and \ref{tab:swnloc}, to a correlation matrix, which also presents variable distribution and scatter plots. This is presented in Figure~\ref{fig:image}.

\section{Conclusions}\label{sec:conclusions}

It can be said that the chatting activities of the developers has predictive power on their productivity. The best predictor of productivity was the number of chat messages sent. In line with previous work, three positive sentiment metrics were associated with higher productivity but to our surprise so was a single negative emotion metric.  

After talking with the developers, it was noted that the chat was overwhelmingly used for discussing aspects related to the project being developed. A separate channel for discussions related to leisure existed, and it is not part of the data analysis efforts for this study. Hence, perhaps much of the developer's activity level is shown in the number of messages sent. 

Previous work on sentiment analysis in the software engineering context has noted the differing results based on the tools used to measure sentiment \cite{jongeling2017negative}. In our case, the usage of the Finnish language in the chat logs dictated the usage of lexicon matching and made the analysis more difficult. Thus, future works with more advanced sentiment analysis tools should be performed. 


\begin{acks}
This work has been supported by Academy of Finland grant 298020.  The first author has been supported by Kaute-foundation.
\end{acks}

\bibliographystyle{ACM-Reference-Format}
\bibliography{sample-base}

\end{document}